\documentclass[aps,superscriptaddress,twocolumn]{revtex4}
\usepackage{amsmath}  
\usepackage{amsfonts} 
\usepackage{newtxtext,newtxmath} 
\usepackage{bm} 
\usepackage{graphicx} 
\usepackage{float} 
\usepackage{subfigure}
\usepackage{epsf} 
\usepackage{epstopdf} 
\usepackage{nameref} 
\usepackage[colorlinks,
            citecolor=blue,
            anchorcolor=red,
            menucolor=red,
            linkcolor=red,
            filecolor=red,
            runcolor=red,
            urlcolor=blue,
            frenchlinks=true]{hyperref}
\usepackage{makecell} 
\usepackage{array} 
\usepackage{xcolor} 
\usepackage{ulem} 

\begin{document}

\title{ HESS J1731-347 is likely a Quark Star Based on the Density-Dependent vMIT Bag Model }

\author{Min Ju}\email{jumin@upc.edu.cn}
\affiliation{School of Science, China University of Petroleum (East China), Qingdao 266580, China}

\author{Pengcheng Chu}\email{kyois@126.com}
\affiliation{The Research Center for Theoretical Physics, Science School, Qingdao University of Technology, Qingdao 266033, China}

\author{Xuhao Wu}\email{wuhaoysu@ysu.edu.cn}
\affiliation{Key Laboratory for Microstructural Material Physics of Hebei Province, School of Science, Yanshan University, Qinhuangdao 066004, China}

\author{He Liu}\email{liuhe@qut.edu.cn}
\affiliation{The Research Center for Theoretical Physics, Science School, Qingdao University of Technology, Qingdao 266033, China}

\begin{abstract}
In this study, we extend the MIT bag model by incorporating the vector interaction among quarks and introducing a density-dependent bag pressure.
Then we proceed to investigate the thermodynamic properties of strange quark matter (SQM) and pure up-down quark matter (udQM) in quark stars (QSs).
Our findings demonstrate that the density dependence of bag pressure $B(n_b)$ and the vector interaction $G_V$ among quarks can significantly stiffen the equation of state (EOS) for both SQM and udQM which allows for the description of massive compact stars such as those observed in GW190814 and PSR J0740+6620 as plausible candidates for QSs.
Ultimately, we derived a series of mass-radius relations of QS based on several combinations of ($B_{as}$, $\beta$). 
Our results support the hypothesis that HESS J1731-347 is a quark star.

\end{abstract}

\maketitle

\section{Introduction}
\label{sec:introduction}
According to quantum chromodynamics (QCD) theory, the deconfinement of quark matter is anticipated to occur at high temperatures and high densities.  
At zero or low temperatures, deconfined quark matter may exist in the core of massive compact stars, such as neutron stars (NSs), where it becomes energetically favorable over hadronic matter~\cite{Annala:2019puf}.
Moreover, Bodmer and Witten proposed that ordinary matter, composed of protons and neutrons, may only be meta-stable~\cite{Bodmer:1971we,Witten:1984rs}. 
The true ground state of strongly interacting matter would therefore consist of strange quark matter (SQM), which in turn is composed of deconfined up, down, and strange quarks. 
If this is true, once the core of the star transitions to the quark phase, the entire star will undergo a conversion into what is known as a quark star (QS) in a short time~\cite{Olinto:1986je}. 

During the past decades, many works have been done on phenomenological models of strongly interacting matter have been done during the past decades~\cite{Lattimer:2006xb,Ozel:2016oaf,Steiner:2014pda}, which have traditionally involved a degree of model dependence.
From recent astronomical observations, massive pulsars~\cite{Antoniadis:2013pzd,NANOGrav:2017wvv} (whose star mass is larger than 2 $M_{\odot}$) have been detected, which implies that the EOS of the star matter should be very stiff.
In 2021, the mass of PSR J0740 + 6620 which, as the most massive precisely observed pulsar, has been updated to 2.08 $\pm$ 0.07 $M_{\odot}$~\cite{Fonseca:2021wxt,Miller:2021qha,Riley:2021pdl}. 
In 2020, the LIGO/Virgo Collaborations declared that the mass of the secondary component in the newly discovered compact binary merger GW190814~\cite{LIGOScientific:2020zkf} could range from 2.50 - 2.67 $M_{\odot}$ at a 90$\%$ credible level. This imposes stringent constraints on EOS of strongly interacting matter, especially when considering the secondary component of GW190814 as a candidate for a compact star. 
The circumferential radius and gravitational mass of the compact star in 4U 1702-429 are estimated as $R = 12.4 \pm 0.4$ km and $M = 1.9 \pm 0.3~ M_{\odot}$ at 68\% credible level, respectively \cite{Nattila:2017wtj}. 
In a very recent study from Ref. \cite{Doro:2022}, the authors provide estimates for the radius and mass of the central compact object within the supernova remnant HESS J1731-347 as $R = 10.4^{+0.86}_{-0.78}$ km and $M = 0.77^{+0.20}_{-0.17}~M_{\odot}$, based on Gaia observations.
These estimates imply that this object is either the lightest known neutron star or a quark star.
Compact stars typically encompass neutron star (NS), quark star (QS), and hybrid star (HS). 
We can not rule out any of them on the basis of our current observational constraints.  
If one considers the supermassive compact stars as QSs, the observations may rule out some of the conventional phenomenological models of quark matter, whereas there still exist some other models which are able to reproduce massive QSs~\cite{Alford:2002rj,Ippolito:2007hn,Paoli:2010kc,Chu:2021aaz,Chu:2023von}. 
The possible existence of QSs is still one of the most important fields of modern particle physics and astrophysics~\cite{Herzog:2011sn,Ivanenko:1969gs}.

Some models have been used to approach the Bodmer and Witten conjecture, the first of them being the original MIT bag model~\cite{Chodos:1974je,Chodos:1974pn}.
MIT bag model has been extensively developed to describe strange matter inside strange stars in~\cite{Farhi:1984qu}.
More sophisticated treatments for SQM, based on the Nambu-Jona-Lasinio ~\cite{Nambu:1961tp,Nambu:1961fr,Menezes:2003pa} and the quark-mass density dependent~\cite{Fowler:1981rp,Chu:2022ofc} models have also been used to this same purpose.
Regarding the original MIT bag model, it was shown it is not able to reproduce massive stars. Some modifications have been made to address this limitation, including the incorporation of a repulsive interaction~\cite{Lopes:2020btp,Klahn:2015mfa,Lopes:2020dvs}.
This modification aims to better describe the properties of massive NS detected in the last decade or so, and it is referred to as the vector MIT (vMIT) bag model.
It is well known that the MIT bag model is characterized by the bag pressure $B$ which is actually the energy density difference between the perturbative vacuum and the true vacuum~\cite{Burgio:2001mk, Burgio:2002sn}. 
From Refs.~\cite{Baym:2017whm, Buballa:2003qv}, the bag constant, which plays a pivotal role in providing the confinement of quarks at low density, is set as $B^{1/4} = 100 \sim 300 $ MeV ($B = 13.0 \sim 1054.2$ MeV fm$^{-3}$), while $B^{1/4} \sim 210$ MeV ($B \sim 253$ MeV fm$^{-3}$) is predicted by matching the EOS of MIT bag model to Lattice QCD (LQCD) calculations~\cite{Benhar:2004gq,Satz:1982fu}. 
Moreover, the bag pressure is also constrained by the tidal deformability from GW170817 observation as $B^{1/4} = 134.1 \sim 141.4 $ MeV ( $B=42.1 \sim 52.0$ MeV fm$^{-3}$) and $B^{1/4} = 126.1 \sim 141.4$ MeV ( $B=32.9 \sim 52.0$ MeV fm$^{-3}$) for the low-spin case and high-spin case within MIT bag model, respectively~\cite{Zhou:2017pha}.
Ref.~\cite{Aziz:2019rgf} employed altogether 20 compact star candidates to constraint the values of $B$, gives an interesting result on the range of the bag constant as 41.58 MeV fm$^{-3}$ $<$ $B$ $<$ 319.31 MeV fm$^{-3}$.
Regarding the bag pressure, there still exists considerable uncertainty due to its model-dependent nature. 
One way to make progress in this situation is to make the models more realistic. Taking the density dependence of $B$ into account is a step in that direction.
The density dependence of $B$ can be seen analogously to the temperature-dependent $B$, where the net inward pressure $B$ must vanish as the temperature rising which means there is no difference between the true vacuum and perturbation vacuum.

In the present work, we use the vMIT bag model to describe QSs while the effective bag pressure $B$ is considered to have density dependence, following a Gaussian distribution as in Refs.~\cite{Bordbar:2020fqj, Miyatsu:2015kwa}.
The Gaussian function to capture the density dependence of bag pressure invokes the asymptotic behavior of the quarks at high densities, which may significantly affect the the structural properties of QSs.
In our investigation, we aim to extend MIT model's capacity to predict the existence of QSs with large masses that align with current astrophysical constraints by incorporating a density-dependent bag pressure and the vector interaction among quarks.
We also want to study whether PSR J1731-347 is a QS based on the vMIT bag model with a density-dependent bag pressure.  

This paper is organized in the following way.
The framework of vMIT bag model and the details to calculate the properties of QSs are presented in Section~\ref{sec:2}. 
The effects of density-dependent bag pressure are shown and discussed in Section~\ref{sec:results}.
Finally, the conclusions are provided in Section~\ref{sec:Con}.

\section{The theoretical formalism}
\label{sec:2}
\subsection{Vector MIT model}
The MIT bag model is employed to describe pure quark matter where the quarks are confined within the “colorless" region commonly referred to as the “bag" that corresponds to an infinite potential.
A key feature of this model is a specific bag constant which is often taken as a free parameter whose value has a wide variation in literatures.
However, it is already known that the original MIT bag model with constant bag pressure cannot satisfy the observational constraints from massive pulsars in case of QSs~\cite{Lopes:2020btp} unless repulsive or perturbative corrections are included.

In this work, we consider the vMIT bag model by including the vector interaction with a vector meson of mass $m_V$, which can be inferred from the $\omega$ meson from Quantum Hadron Dynamics (QHD)~\cite{Serot:1992ti}.
The Lagrangian density of vMIT bag model is expressed as follows: 
\begin{eqnarray} 
\label{eq:LvMIT}
\mathcal{L_{\text{vMIT}}} &=& \sum_{i=u,d,s}\Big[ \bar{\psi}_{i}\left( i\gamma_{\mu}\partial^{\mu}
	          -m_{i}-g_{iiV}\gamma_{\mu}V^{\mu}\right)\psi_{i} \nonumber\\
            &&  +\frac{1}{2}m_{V}^{2}V_{\mu}V^{\mu}    
	        -B \Big ] \Theta(\bar{\psi}_{i}\psi_{i})  \nonumber\\
	    &&  +\sum_{l=e,\mu}\bar{\psi_{l}}\left( i\gamma_{\mu}\partial^{\mu}
	        -m_{l}\right)\psi_{l}, 
\end{eqnarray}
where $m_i$ is the mass of the quark $i$ of flavor $u,d$ and $s$. Here we use the current quark masses $m_u = m_d = 5.5$ MeV and $m_s = 95$ MeV in our calculations; $\psi_{i}$ is the Dirac quark field, $B$ is the vacuum pressure, and $\Theta(\bar{\psi}_{i}\psi_{i})$ is the Heaviside step function to assure that the quarks are confined within the bag. Leptons ($e$ and $\mu$) are added to account for the $\beta$-equilibrium matter. 
The interaction among quarks is mediated by the massive vector channel $V_{\mu}$.

Regarding the introduction, we utilize density-dependent bag pressure, as outlined in the Refs.~\cite{Burgio:2001mk,Burgio:2002sn}.
The density dependent bag pressure, which assumes finite values $B_{0}$ at $n_b=0$ and $B_{as}$ at asymptotic density, is modeled by a Gaussian distribution as follows
\begin{equation} 
	\label{eq:B}
	B(n_b) = B_0 - (B_0-B_{as})\left[1-\text{exp}(-\beta(\frac{n_b}{n_0})^2) \right],
\end{equation}
where the parameter $\beta$ governs the gradual decrease of $B$ as density increases. $n_0$ is the nuclear matter saturation density and we set $n_0=0.15$ fm$^{-3}$ in our calculation.
We mainly use $B_{0}=257.3$ MeV fm$^{-3}$, which is determined by matching the mass and radius of free nucleons using MIT model as described in Ref.~\cite{Panda:2003zj}. Additionally, $B_{0}=136.6$ MeV fm$^{-3}$ is also used for comparison to discuss the effects from different initial pressures.
It is worth mentioning that $B_{0}$ represents the values at $n_b=0$, which is not very important for QSs since they have a finite surface baryon density~\cite{Zhang:2021qhl}.

Then we can get the eigenvalue energy of quarks and the equation of motion for the $V$ field, respectively,
\begin{eqnarray}
	\epsilon_{i}   &=& \sqrt{m_{i}^{2}+k^{2}}+g_{iiV}V_{0}  \notag \\
    && -2\beta(B_0-B_{as}) \frac{n_b}{3n_0^2}  \exp\big(-\beta(\frac{n_b}{n_0})^2\big),  \\
	m_{V}^{2}V_{0} &=& \sum_{i=u,d,s}g_{iiV}\langle\bar{\psi_{i}}\gamma^{0}\psi_{i}\rangle,
\end{eqnarray}
where the term $\langle\bar{\psi_{i}}\gamma^{0}\psi_{i}\rangle$
can be interpreted as the number density $n_{i}$ for each $i$ flavor of quarks.
The potential of $i$ quark ($i$=$u$, $d$, $s$) is given by 
\begin{eqnarray}
	\mu_{i} 
	&=& \sqrt{{k_{F}^i}^{2}+m_{i}^{2}} +g_{iiV}V_0   \nonumber\\ 
	&-& 2\beta(B_0-B_{as})\frac{n_b}{3n_{0}^{2}}\exp\big(-\beta(\frac{n_b}{n_{0}})^2\big).
\end{eqnarray}
The total energy density and and pressure in quark matter are calculated as follows:
\begin{eqnarray}
	\varepsilon &=& \sum_{i=u,d,s} \varepsilon^i_{\rm{FG}}
	+\sum_{l=e,\mu} \varepsilon^l_{\rm{FG}}  +\frac{1}{2} m_V^2 V_0^2 +B(n_b) ,
	\label{eq:eqp}\\
	P &=& \sum_{i=u,d,s} P^i_{\rm{FG}}
	+\sum_{l=e,\mu} P^l_{\rm{FG}} +\frac{1}{2} m_V^2 V_0^2 -B(n_b)  \nonumber\\
&&      -2\beta(B_0-B_{as})\frac{{n_b}^2}{3n_{0}^{2}}\exp\big(-\beta(\frac{n_b}{n_{0}})^2\big).
	\label{eq:pqp}
\end{eqnarray}
Here $\varepsilon_{\rm{FG}}^{i}$ and $P_{\rm{FG}}^{i}$ denote the Fermi gas contributions of species $i$ with a mass $m_i$ and degeneracy $N_i$,
\begin{eqnarray}
	\varepsilon^i_{\rm{FG}} &=& N_i
	\int_{0}^{k^{i}_{F}} \frac{d^{3}k}{(2\pi)^3} \sqrt{k^2+m_i^2},
	\label{eq:efg}   \\
	P^i_{\rm{FG}} &=& \frac{N_i}{3}
	\int_{0}^{k^{i}_{F}} \frac{d^{3}k}{(2\pi)^3} \frac{k^2}{\sqrt{k^2+m_i^2}} .
	\label{eq:pfg}
\end{eqnarray}
For the $i$-th flavor of quarks, the degeneracy $N_i = 6$, while for the leptons $N_i = 2$. 
The energy per baryon can be obtained by $E/A=\varepsilon /n_b$.

In vMIT bag model, two important quantities $X_V$ and $G_V$, as suggested in Ref.~\cite{Lopes:2020btp}, are defined as,
\begin{equation}
	X_V=\frac{g_{ssV}}{g_{uuV}},\quad G_V=(\frac{g_{uuV}}{m_V})^2.
\end{equation}
$X_V$ is related to the strength of the vector field with the $s$ quark in relation to $u$ and $d$ quarks. 
$X_V=1.0$ is commonly used in Refs.~\cite{Menezes:2014aka,Lopes:2020btp,Lopes:2019shs} which can support a relatively massive QS.
$G_V$ is related to the absolute strength of the vector field which is crucial for high density EOS of QSs, and there are very few studies trying to constrain its absolute value \cite{Restrepo:2014fna}. Most of the models just consider it as a free parameter \cite{Menezes:2014aka,Shao:2012tu}. 
The commonly used values of $G_V$  are in the range of 0 - 0.5 fm$^{2}$, while some larger values are also used in the study \cite{Ogihara:2019rds, Franzon:2016urz}. In the present work, we use $G_V$ = 0.3 fm$^2$ as a conservative choice.

\subsection{Quark Star}
In the present work, $B_{as}$ and $\beta$ are treated as free parameters to investigate their influences on the properties of QSs. The Bodmer-Written conjecture posits that at high density, the presence of strange quark turns the deconfined quark matter into the true ground state of strong interaction by reducing its binding energy. 
This conjecture can be succinctly summarized as ensuring the binding energy of SQM is lower than that of iron, while the binding energy of iron is lower than that of udQM.
By imposing these constraints, we can establish the stability window for the parameters 
$B_{as}$ and $\beta$ within the framework of the vMIT model. This allows us to identify the ranges of these parameters that ensure stable quark star configurations.

QS comprises leptons (here we consider $e$ and $\mu$) and quarks because of the necessary charge neutrality and $\beta$ equilibrium given by
\begin{eqnarray}
	n_e+n_{\mu} &=& \frac{1}{3}(2n_u-n_d-n_s),  \\
	\mu_{d}=\mu_{s} &=& \mu_{u}+\mu_{e}; \quad \mu_e = \mu_{\mu}.
\end{eqnarray}
To calculate the structures and tidal deformability of QSs, we briefly introduce the relevant equations. 
To derive the structural properties of QSs, one must solve the Tolman-Oppenheimer-Volkov (TOV) equations,
\begin{equation}
	\frac{dP}{dr} = -\frac{[P(r)+\varepsilon(r)][M(r)+4\pi r^3P(r)]}{r(r+2GM(r))},
\end{equation}
\begin{equation}
	\frac{dM(r)}{dr} = 4\pi r^2 \varepsilon(r).
\end{equation}
Dimensionless tidal deformability is denoted as
\begin{equation}
	\Lambda	= \frac{2}{3}k_2C^{-5},
\end{equation}
where $C=M/R$ is the compactness and $k_2$ is the gravitational Love number.
Then we could get the maximum mass of QSs numerically which is closely related to the speed of sound.
$c_s^2=\partial P(r)/\partial \varepsilon(r)$ denotes the squared speed of sound which could provide valuable insights into the microscopic description of dense matter.
\begin{figure}[htbp]
\hspace{-1.0cm}
\includegraphics[width=0.9\linewidth]{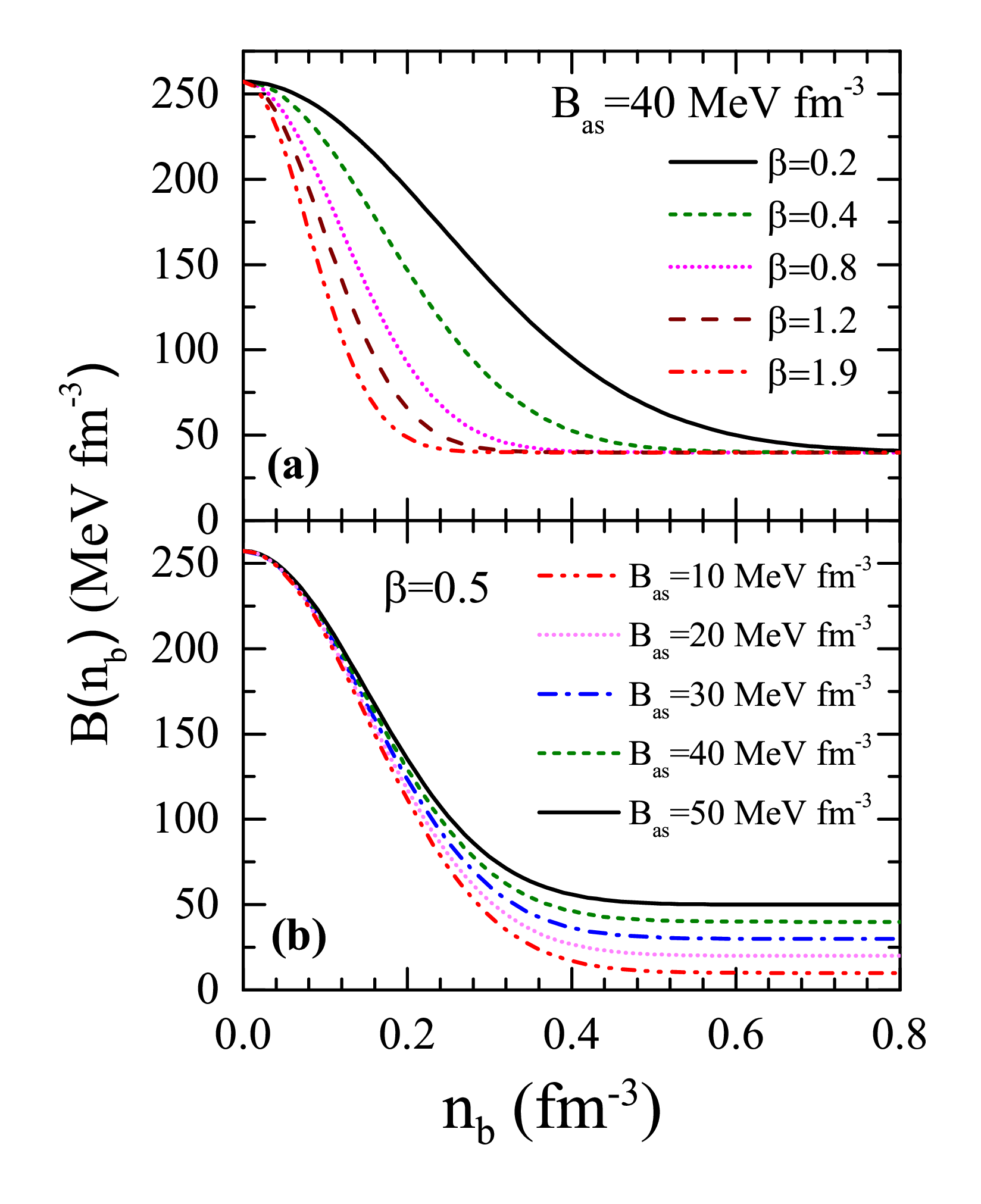}\\
\caption{ (a) Variation of bag pressure $B(n_b)$ with respect to density $n_b$ for different $\beta$ by fixing $B_{as}=40$ MeV fm$^{-3}$.
(b) Variation of bag pressure with density for different values of $B_{as}$ fixing $\beta=0.5$ and $B_{0}=257.3$ MeV fm$^{-3}$ for both panels.}
\label{fig:1B}
\end{figure}
%
%
\section{Results and discussion}
\label{sec:results}
\begin{figure}
\hspace{-27pt}
\includegraphics[width=1.10\linewidth]{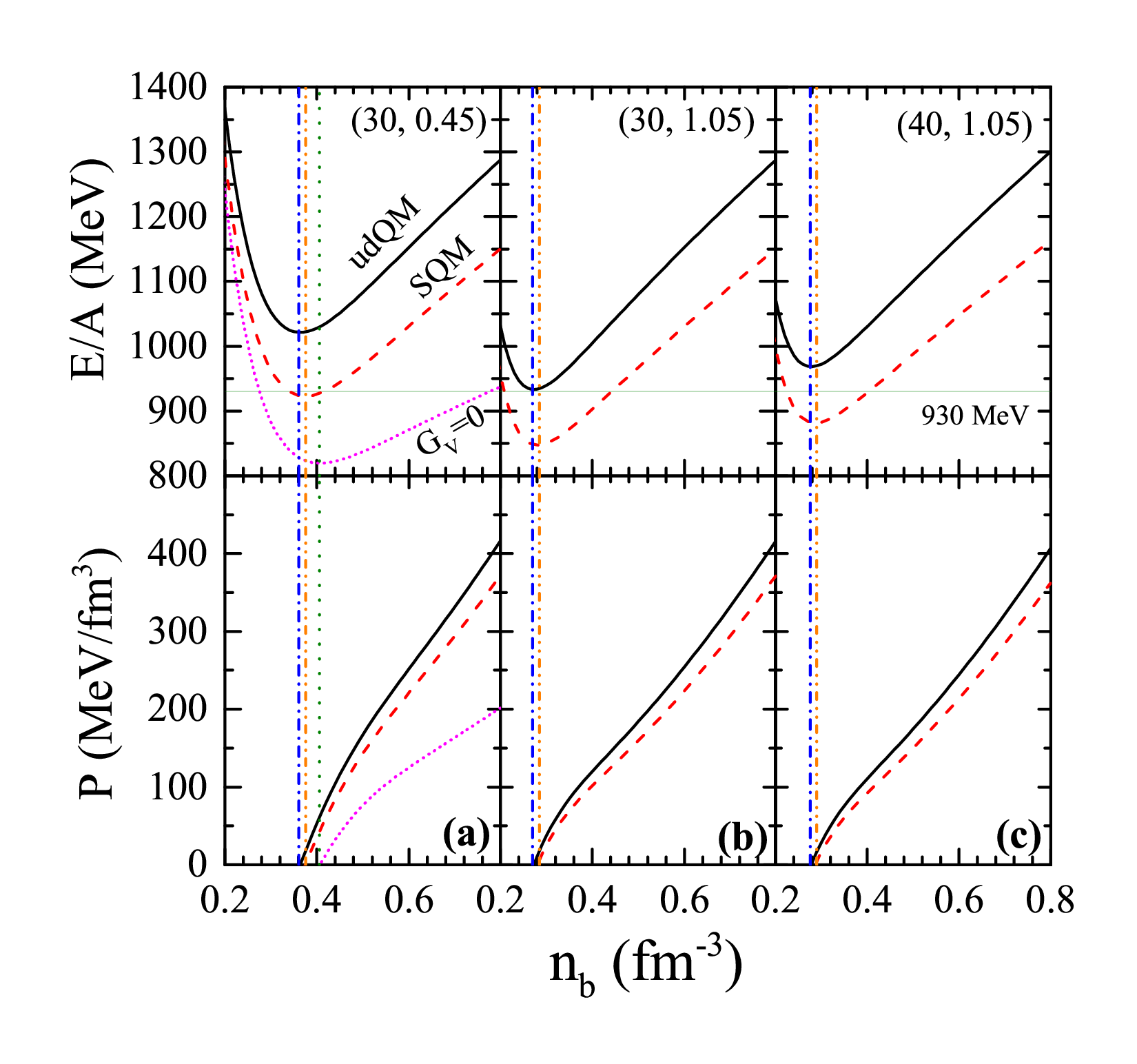}
\caption{ Energy per baryon and pressure as functions of baryon number density based on the vMIT bag model with density dependent bag pressure for different parameters combinations ($B_{as}$, $\beta$) when we set $B_{0}=257.3$ MeV fm$^{-3}$ and $G_{V}=0.3$ fm$^{2}$. The black solid line represents the results for udQM while the red dashed line represents for SQM. The vertical dash-dot and the dash-dot-dot lines correspond to the lowest energy point of the udQM and SQM, respectively.
The purple dotted line in (a) is the result for SQM in the case $G_{V}=0$ when using (30, 0.45) and the same $B_{0}$.}   
\label{fig:2PE-nb}
\end{figure}
\begin{figure*}
    \hspace{-30pt}
    \includegraphics[width=1.0\linewidth]{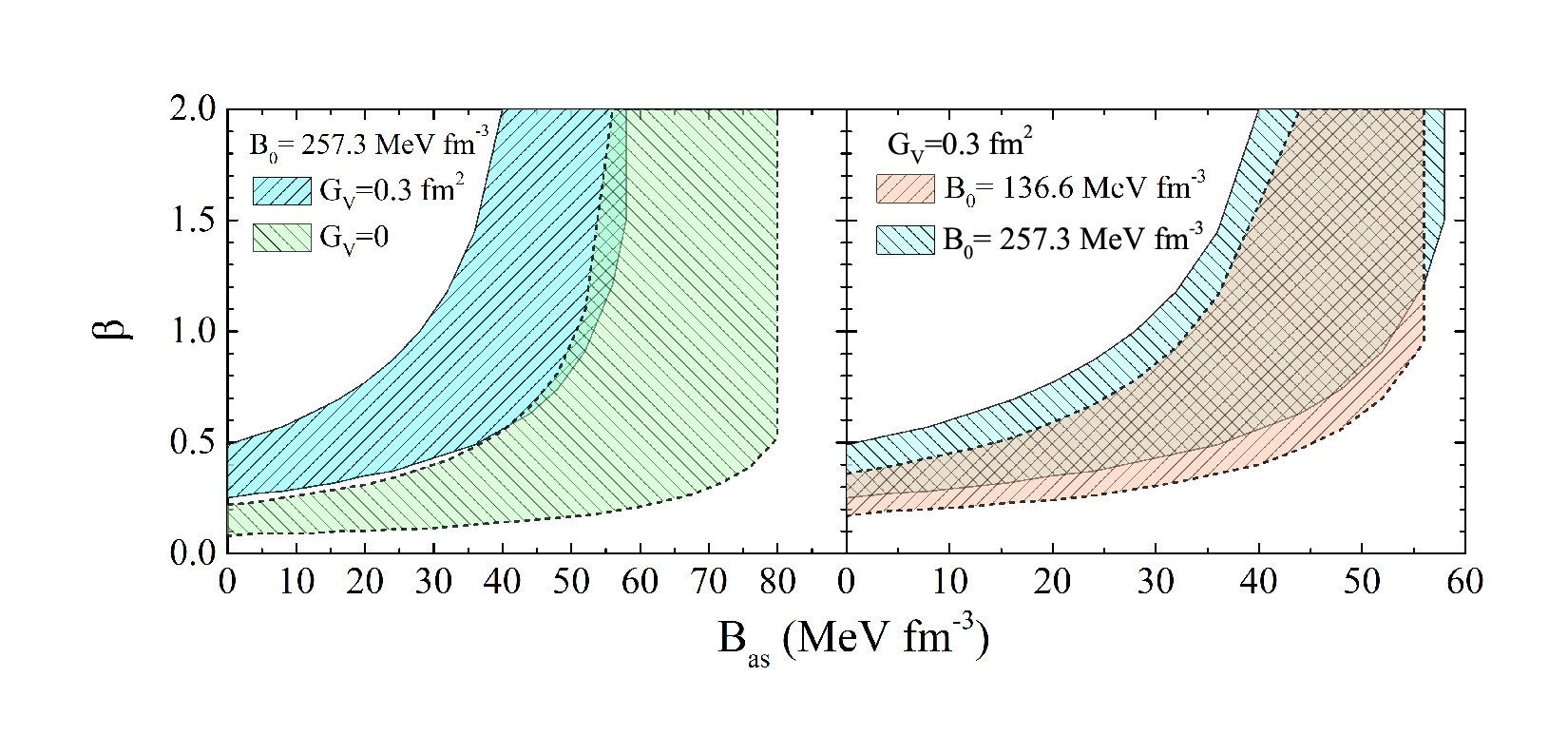}
    \caption{ The stable parameter space of $B_{as}$ and $\beta$ for two choices of $G_V$ (0 and 0.3 fm$^2$) and $B_0$ (257.3 and 136.6 MeV fm$^{-3}$).  }  
    \label{fig:3B}
\end{figure*}
\begin{figure}
	\centering
        \hspace{-30pt}
	\includegraphics[width=1.1\linewidth]{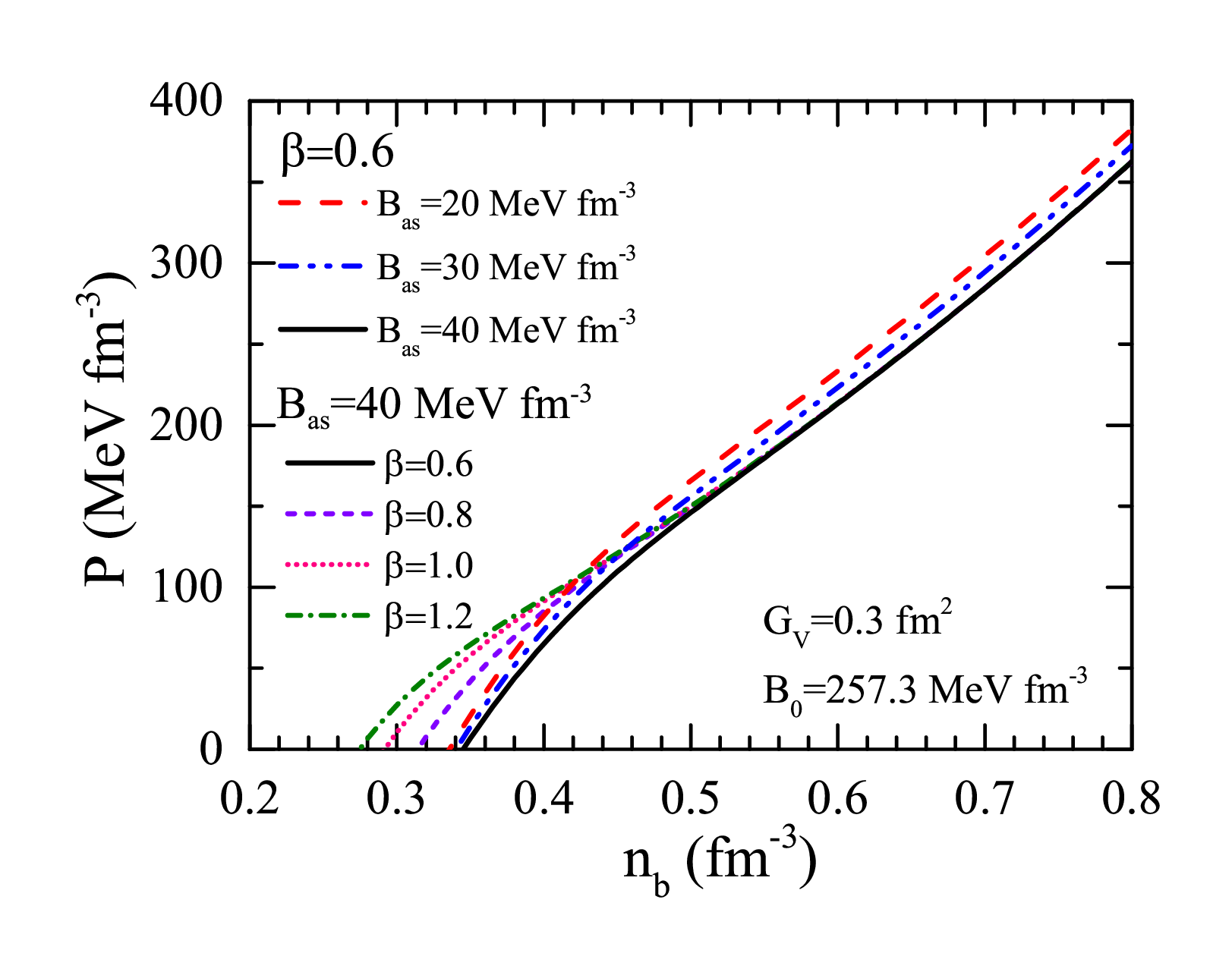}
	\caption{ EOSs of QS matter with density-dependent bag pressure for different parameter combinations ($B_{as}$, $\beta$). }
	\label{fig:4P-nb}
\end{figure}
\begin{figure*}
\hspace{-40pt}
\includegraphics[ width=0.9\linewidth]{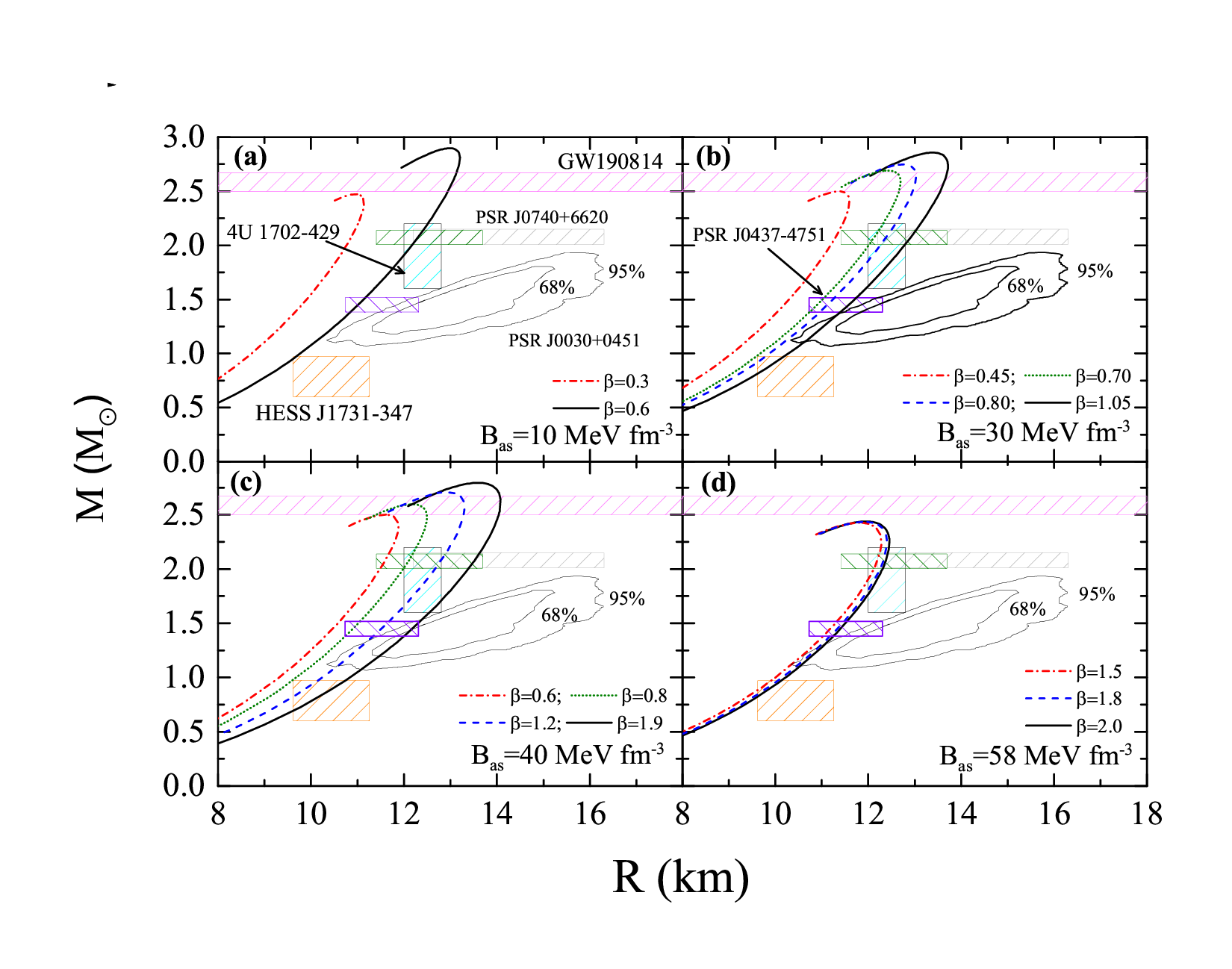}%
\caption{  Mass-radius relations of QSs based on vMIT bag model with varying parameters $B_{as}$ and $\beta$ within the stability window. 
All the results are obtained under the conditions $B_{0}=257.3$ MeV fm$^{-3}$ and $G_{V}=0.3$ fm$^{2}$.
The orange shaded area is the constraint from HESS J1731-347~\cite{Doro:2022,Hong:2024sey} with 68\% credible level, while the constraints from 4U 1702-429~\cite{Nattila:2017wtj} and PSR J0437-4751~\cite{Choudhury:2024xbk} are also shown for comparison.
Observational limits imposed from PSR J0740+6620 on maximum mass and radius~\cite{Miller:2021qha,Riley:2021pdl} are also indicated.
The simultaneous measurement of the mass and radius for PSR J0030+0451 by NICER with 68\% and 95\% confidence intervals are also shown~\cite{Miller:2019cac}.
The mass constraint from GW190814~\cite{LIGOScientific:2020zkf} is depicted by the pink horizontal bar. 
}
\label{fig:5mr}
\end{figure*}
%
%
 In this section, we explore the properties of QSs using the vMIT bag model, with a Gaussian distribution for the bag pressure $B(n_{b})$ following Eq.~(\ref{eq:B}). 
 There are mainly two key parameters ($B_{as}$, $\beta$) govern the variations of bag pressure.
 We start with concentrating on the density dependent scenario of bag pressure.
 As mentioned in the previous section, $B_0 = 257.3$ MeV fm$^{-3}$ is mainly used. 
 In Fig.~\ref{fig:1B}(a), we show the bag pressure as a function of $n_b$ at $B_{as}$ = 40 MeV fm$^{-3}$ with $\beta = 0.2-1.9$. 
 As the parameter $\beta$ increases, the asymptotic freedom gained by $B(n_b)$ is more earlier. 
 Moreover, when $\beta$ exceeds 1.2, the influence of $\beta$ becomes less significant (specifically after $n_b$ surpasses 0.3 fm$^{-3}$). 
 For instance, when $\beta$ = 1.9, $B(n_b)$ approaches to the asymptotic freedom at $n_b \approx 0.3$ fm$^{-3}$, which indicates a very strong density dependence of the bag pressure with $\beta$ in vMIT bag model.
 The Fig.~\ref{fig:1B}(b) shows the results from $B_{as}$= 10 - 50 MeV fm$^{-3}$ when fixing $\beta=0.5$.
 We found that the decrease in $B(n_b)$ occurs more rapidly as the value of $B_{as}$ decreases. $B_{as}$ has a pronounced effect on $B(n_b)$, particularly when the baryon density $n_b$ exceeds 0.3 fm$^{-3}$.
 In summary, parameters ($B_{as}$, $\beta$) play significant roles at different density ranges in modulating the behavior of $B(n_b)$. 

In Fig.~\ref{fig:2PE-nb}, we present the energy per baryon ($E/A$) and the pressure ($P$) for both SQM and udQM as a function of baryon number density ($n_b$) with the vector coupling $G_{V}=0.3$ fm$^{2}$. 
The black solid line represents the results for udQM while the red dashed lines represent for SQM.
According to the Bodmer-Witten conjecture, if the true ground state of strongly interacting matter consists of SQM, the absolutely stable condition should be considered, which requires the minimum energy per baryon of SQM or udQM should be less or larger than the energy per baryon of the known most stable nuclei $E/A(^{56}\text{Fe})=930$ MeV.
From Fig.~\ref{fig:2PE-nb}, one can see that the baryon density where the minimum energy per baryon occurs corresponds exactly to the zero-pressure point (which can also be considered as the surface density of QSs) density across all cases, ensuring thermodynamic self-consistency.
In Fig.~\ref{fig:2PE-nb} (a), the minimum energy per baryon for SQM is approximately 930 MeV, which is attributed to the fact that $\beta=0.45$ represents the lower limit in the case of $B_{as}=30$ MeV~fm$^{-3}$ which can be understood in the Fig.~\ref{fig:3B} that shows the absolutely stable window of ($B_{as}$, $\beta$). 
As depicted in Fig.~\ref{fig:2PE-nb} (b), the minimum energy per baryon for udQM is approximately 930 MeV which means the upper limit of $\beta$ for $B_{as}=30$ MeV~fm$^{-3}$ is 1.05. 
Comparing the left two panels, one can find that the minimum energy per baryon of both SQM and udQM decreases with $\beta$, which indicates that one can increase $\beta$ from the lower limit case (when the minimum energy of the SQM increases to 930 MeV) to upper limit case (when the minimum energy per baryon of the udQM decreases to 930 MeV). 
Furthermore, we find that the minimum energy per baryon of SQM with $G_{V}=0$ is obviously less than the case when $G_{V}=0.3$ fm$^{2}$ from Fig.~\ref{fig:2PE-nb} (a), which implies that the presence of vector interaction among quarks leads to a significant stiffening of EOS.
Comparing the right two panels, one can also find that the energy per baryon for both SQM and udQM increases with $B_{as}$ when $\beta$ is fixed.
Compared to $\beta$, $B_{as}$ has little effect on the surface density of QSs, as can be clearly seen in Fig. \ref{fig:4P-nb}. 

The stability window for ($B_{as}$, $\beta$), based on the absolutely stable condition in two scenarios, different $G_V$ and different $B_0$, is shown in Fig. \ref{fig:3B}.
From the left panel, one can see that the addition of vector interaction shifts the stability window of ($B_{as}$, $\beta$) to the left and upward, while also narrowing the stability region. This indicates that the strong interaction among quarks will constrain the bag pressure to some extent.
As the value of $B_0$ transitions from 257.3 to 136.6 MeV fm$^{-3}$, the stability region is significantly reduced.
Furthermore, the results show the existence of a distinct right boundary of the `stability window' in the parameter space ($B_{as}$, $\beta$).
For example, in the case with $B_{0}=257.3$ MeV fm$^{-3}$ ($B_{0}=136.6$ MeV fm$^{-3}$) and $G_V$=0.3 fm$^2$, the right boundary of the stability window, namely the maximum value of $B_{as}$, is 58 MeV fm$^{-3}$ (56 MeV fm$^{-3}$).

In Fig.~\ref{fig:4P-nb}, we present the EOSs of QSs by systematically varying $B_{as}$ and $\beta$, which reside within the stability window. We examine the structural properties of QSs by varying one parameter at a time and keeping the other one fixed.
We find that $B_{as}$ does not affect the surface density of QSs much, while the surface density obviously decreases with increasing $\beta$.
One can see that both $B_{as}$ and $\beta$ predominantly affect the EOS at the lower densities. Compared to \(B_{as}\), \(\beta\) exhibits a more pronounced effect on the EOS.
It can be speculated that within the stability window, $\beta$ might be the key parameter in determining the maximum mass of QS, implying that variations in \(\beta\) could have a significant effect on the structural properties of QSs. 
Combining with Fig. \ref{fig:1B}, we find that as $\beta$ increases, the density at which the curve of $B(n_b)$ approaches the asymptotic freedom, gets closer to the surface density, meaning that it approaches a constant bag pressure. 
For example, for $\beta=1.9$ and $B_{as}$ = 40 MeV fm$^{-3}$, the density at which $B(n_b)$ decreases to $B_{as}$ is around 0.3 fm$^{-3}$ as depicted in Fig. \ref{fig:1B}, while the surface density is 0.275 fm$^{-3}$.
This is why we do not recommend larger values of $\beta$. In the following calculations, we use $\beta = 2.0$ as the upper limit value.

\setcellgapes{3pt} 
\makegapedcells 
\setcellgapes{3pt} 
\setlength{\tabcolsep}{3.0mm}   
\begin{table*}
    \centering
    \begin{tabular}{cccccccccccc}
\hline
      $G_{V}$ & $B_0$ & $B_{as}$ & $\beta$ & $n_s$  & $n_c$ & $\epsilon_{c}$ & $M_{\text{max}}$ & $R_{\text{max}}$ & $R_{1.4}$ & $\Lambda_{1.4}$  \\
       (fm$^2$)  & (MeV fm$^{-3}$) & (MeV fm$^{-3}$) &  &  (fm$^{-3}$) &  (fm$^{-3}$) & (MeV fm$^{-3}$)&   ($M_{\odot}$) &  (km) &  (km) &    \\[0.3ex]
\hline
0.3  & 257.3 & 10 & 0.30  & 0.52 & 0.91 & 820.7 & 2.47 & 10.96 & 9.73   & 30.9 \\
     &       & 10 & 0.60  & 0.40 & 0.70 & 735.3 & 2.90 & 12.96 & 10.77  & 57.6 \\
     \hline
0.3  & 257.3 & 20 & 0.35  & 0.49 & 0.89 & 1062.9 & 2.47 & 11.07 & 9.85   & 28.6 \\
     &       & 20 & 0.60  & 0.40 & 0.73 & 795.9  & 2.76 & 12.53 & 10.69  & 48.2 \\
     &       & 20 & 0.70  & 0.38 & 0.73 & 795.9  & 2.84 & 12.90 & 10.97  & 55.0 \\
     &       & 20 & 0.75  & 0.37 & 0.69 & 745.3  & 2.88 & 13.13 & 11.10  & 57.1 \\
     \hline
     &       & 30 & 0.45  & 0.44 & 0.83 & 975.1 & 2.50 & 11.38 & 10.09  &  39.0 \\    
     &       & 30 & 0.60  & 0.40 & 0.77 & 846.1 & 2.63 & 12.06 & 10.54  &  45.2 \\
0.3  & 257.3 & 30 & 0.70  & 0.38 & 0.76 & 856.1 & 2.69 & 12.39 & 10.78  &  48.6 \\
     &       & 30 & 0.80  & 0.36 & 0.72 & 795.8 & 2.75 & 12.73 & 11.00  &  55.4 \\
     &       & 30 & 1.05  & 0.32 & 0.68 & 734.9 & 2.86 & 13.37 & 11.46  &  64.3 \\
      \hline
     &       & 40 & 0.60  & 0.39 & 0.81 & 945.7 & 2.50 & 11.61 & 10.35  &  40.1 \\
     &       & 40 & 0.70  & 0.37 & 0.79 & 906.0 & 2.56 & 11.91 & 10.61  &  44.9 \\
0.3  & 257.3 & 40 & 0.80  & 0.35 & 0.76 & 866.1 & 2.60 & 12.18 & 10.78  &  50.3 \\
     &       & 40 & 1.20  & 0.30 & 0.69 & 765.3 & 2.71 & 12.94 & 11.39  &  65.8 \\
     &       & 40 & 1.90  & 0.25 & 0.67 & 724.4 & 2.79 & 13.57 & 12.05  &  91.9 \\
    \hline
0.3  & 257.3 & 58 & 1.50  & 0.28 & 0.81 & 959.7 & 2.43 & 11.81 & 11.24  &  66.1 \\
     &       & 58 & 1.80  & 0.26 & 0.80 & 959.7 & 2.43 & 11.92 & 11.26  &  68.7 \\
     &       & 58 & 1.90  & 0.26 & 0.80 & 959.7 & 2.44 & 11.94 & 11.27  &  69.0 \\
     &       & 58 & 2.00  & 0.26 & 0.80 & 959.7 & 2.44 & 11.94 & 11.28  &  72.9 \\
     \hline
0.3  & 136.6 & 30 & 0.80 & 0.30 & 0.76 & 856.1 & 2.82 & 13.08 & 11.51  &  72.5 \\
     \hline
    \end{tabular}
    \caption{ QS main properties for different values of $B_{as}$ and $\beta$. $n_s$ represents the surface density of the QS, while $n_c$ and $\epsilon_c$ denote the central density and the central energy density, respectively.}
    \label{tab:QS}
\end{table*}
\begin{figure}[htbp]
\hspace{-30pt}
\includegraphics[width=1.0\linewidth]{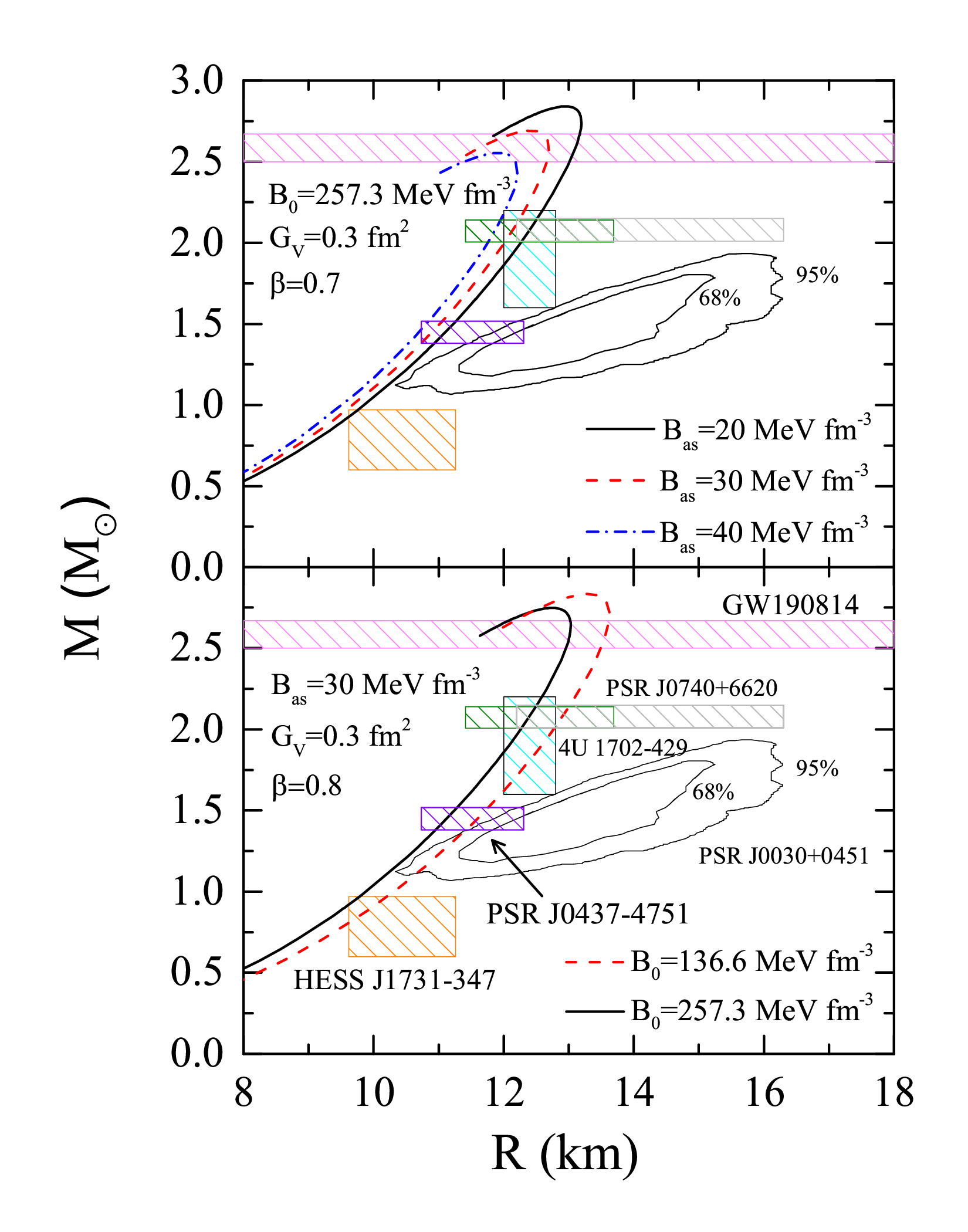}
\caption{ Mass-radius relations under different values of $B_{as}$ and $B_0$. }
\label{fig:6mr}
\end{figure}
\begin{figure*}[htbp]
\hspace{-30pt}
\includegraphics[width=0.8\linewidth]{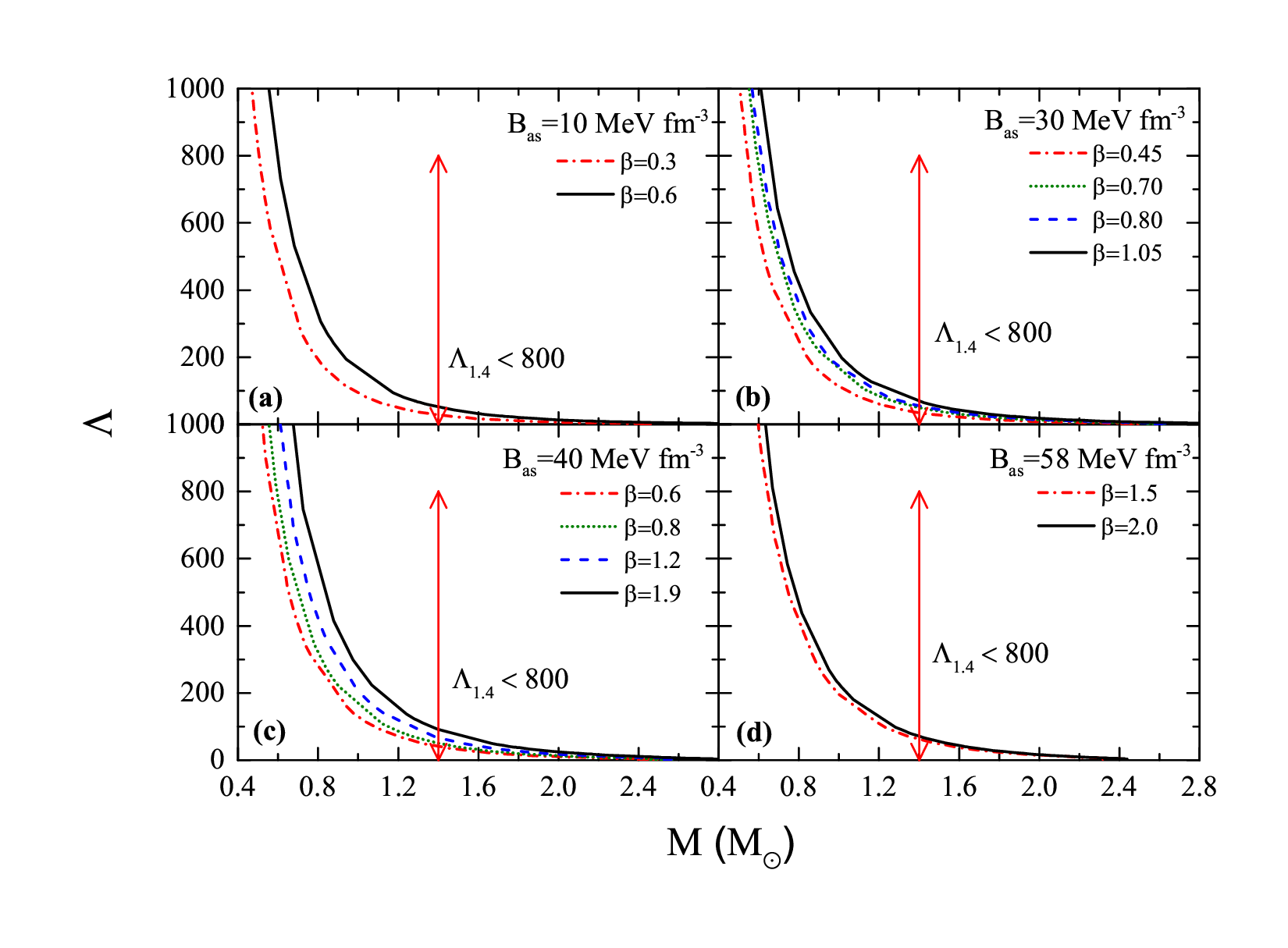}%
\caption{ Dimensionless tidal deformability as a function of the gravitational mass for different parameter combinations ($B_{as}$, $\beta$).
All the results are under the case when $B_{0}=257.3$ MeV fm$^{-3}$ and $G_{V}=0.3$ fm$^{2}$.
The red vertical line is the constraint $\Lambda_{1.4}<800$ at the 90\% confidence level from GW170817~\cite{LIGOScientific:2017vwq}. }
\label{fig:7Lambda}
\end{figure*}
With the stability window in mind, we can study the masses and radius of QSs for every combination of parameters ($B_{as}$, $\beta$).
It should be noted that all EOSs used as input data to solve the TOV equations satisfy the causality condition, $c_s^2 < 1$, ensuring that the speed of sound does not exceed the speed of light.
In Fig.~\ref{fig:5mr}, we present a series of EOSs based on the vMIT bag model under the case $B_0=257.3$ MeV fm$^{-3}$ and $G_{V} = 0.3$ fm$^{2}$. 
The red dash-dot line represents the results corresponding to the lower limit of $\beta$, while the black solid line corresponds to the upper limit of $\beta$ and the green dotted line reflects the minimum value of $\beta$ which could satisfy the constraints from HESS J1731-347. 
For example, in Fig.~\ref{fig:5mr} (a), $\beta=0.3$ (red dash-dot line) is the lower limit which implies the energy per baryon of udQM decreases to 930 MeV, while $\beta=0.6$ (black solid line) is the upper limit, which corresponds that the energy per baryon of SQM increases to 930 MeV. 
We find that not all the parameters recommended based on the absolute stable condition can satisfy the constraint from HESS J1731-347. 
For example, the M-R curve cannot pass through the region of HESS J1731-347 until $\beta$ exceeds 0.7 for the case $B_{as}=30$ MeV fm$^{-3}$.
For a moderate value of \( B_{as} \), there is considerable parameter space in selecting \(\beta\) to support the hypothesis that HESS J1731-347 is a QS. Similar hypotheses have been reported in Refs.~\cite{DiClemente:2022wqp,Restrepo:2022wqn,Horvath:2023uwl,Oikonomou:2023otn,Rather:2023tly}.
Additionally, we find that the M-R curve satisfies the constraints from 4U 1702-429, PSR J0437-4751 and PSR J0740+6620, provided that it intersects with the HESS J1731-347 constraint. 
In other words, if HESS J1731-347 is confirmed to be a QS, then 4U 1702-429, PSR J0437-4751, and PSR J0740+6620 are also likely to be QSs. It is important to emphasize that this conclusion is model-dependent and requires validation through additional models.
The maximum mass of QSs increases with $\beta$ and is larger than $2.4~M_{\odot}$, which indicates that we can employ the vMIT bag model with density-dependent bag pressure to describe extremely massive QSs by varying $B_{as}$ and $\beta$.
Moreover, we demonstrate that the mass-radius relation predicted by the vMIT bag model with density-dependent bag pressure intersects with the observation-constrained mass region of GW190814, which indicates that the secondary component of GW190814 is probably to be a QS.  

The main properties of QSs corresponding to the Fig. ~\ref{fig:5mr} are summarized in Tab.~\ref{tab:QS}. 
One can observe that the maximum mass of QSs increases with increasing $\beta$ and decreasing $B_{as}$.
This can be explained from the Fig.~\ref{fig:1B}, $B(n_b)$ descends more steeply with larger $\beta$ and smaller $B_{as}$. The EOS becomes more stiffer with faster declined $B(n_b)$. 
The maximum value of QSs ($M_{\text{max}} \approx 2.90~M_{\odot}$) is achieved under the conditions $G_{V} = 0.3$ fm$^{2}$, $B_{as}=10$ MeV fm$^{-3}$ and $\beta=0.60$. 
Based on the vMIT bag model with density-dependent bag pressure, this configuration supports a QS with a mass heavier than approximately $2.40~M_{\odot}$. 
Furthermore, the results also indicate that the central baryon density of the maximum mass of QSs decreases with the increment of the star mass, which is consistent with the conclusion from Ref.~\cite{Chu:2023von}.
A smaller $B_0$ leads to a lower surface density and a higher central baryon density, which explains why it supports a more massive QS as depicted in Fig.~\ref{fig:6mr}.
When $B_{as}$ reaches its upper limit of 58 MeV fm$^{-3}$, the structural properties of QSs show only minor variations.

We explored the effects of varying $B_{as}$ and $B_0$ on the mass-radius relations in Fig.~\ref{fig:6mr}.
We observe that as $B_{as}$ or $B_0$ decreases, the maximum mass of QSs increases.
This can be easily understood from Eq.~(\ref{eq:pqp}), where the EOS becomes stiffer with a smaller $B_0$ which contributes to the formation of more massive QSs.  
In the case of $\beta=0.7$, one can find that it does not satisfy the constraint imposed by HESS J1731-347 but still within the constraints from 4U 1702-429 and PSR J0437-4751 when $B_{as}>30$ MeV fm$^{-3}$.
It is important to highlight that the maximum mass of QSs is lower than 1.3 $M_{\odot}$ when employing the original MIT bag model with a constant bag pressure $B = B_0$ = 257.3 MeV fm$^{-3}$.  
The density dependence of bag pressure can significantly stiffen the EOS of QSs.
For instance, with $B_{0}=$ 257.3 MeV fm$^{-3}$, the maximum mass of QSs ranges from  1.7 to 2.0 $M_{\odot}$ when using various combinations of parameters ($B_{as}$, $\beta$) within the stability window. 
Furthermore, the addition of vector interaction among quarks can increase the maximum mass of QSs by $\sim$ 0.5 $M_{\odot}$ as the interaction strength varies from 0 to 0.3 fm$^2$. 

\begin{figure*}
\centering
\hspace{-45pt}
\includegraphics[width=1.0\linewidth]{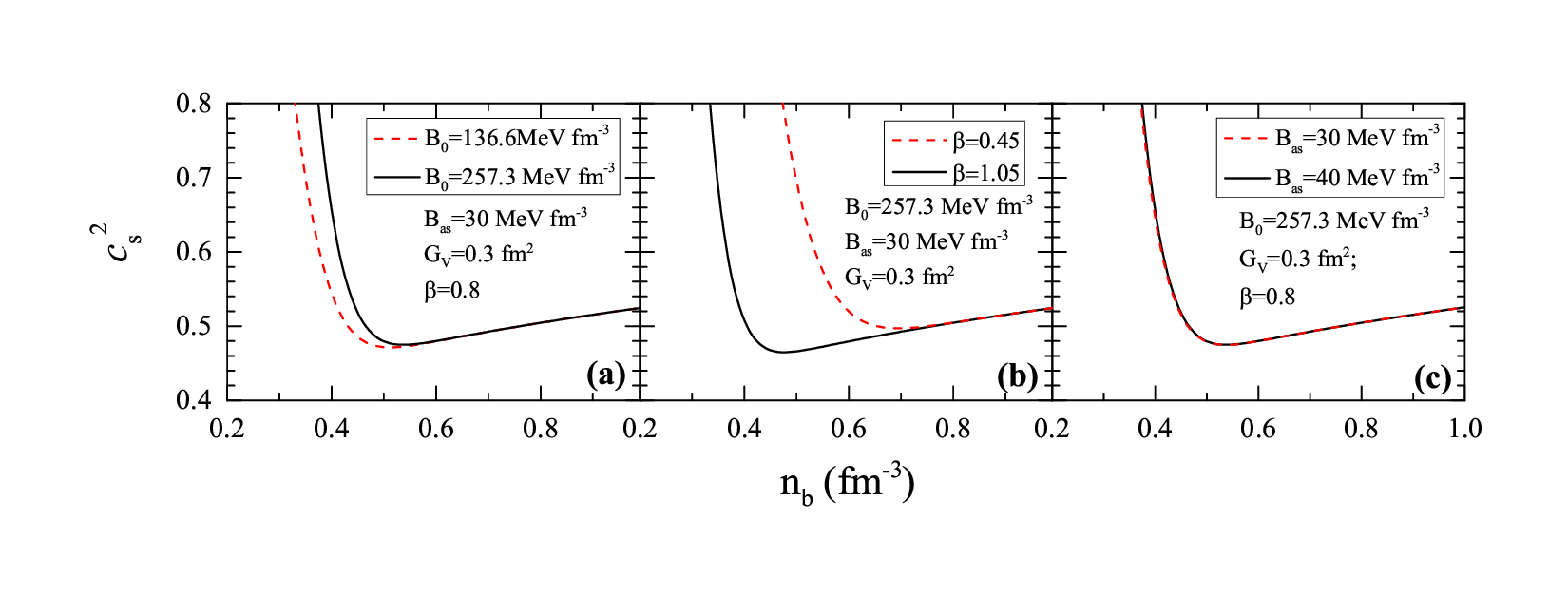}
\caption{ The speed of sound as a function of the baryon number density.}
\label{fig:8vs2}
\end{figure*}
\begin{figure}
\centering
\hspace{-40pt}
\includegraphics[width=1.1\linewidth]{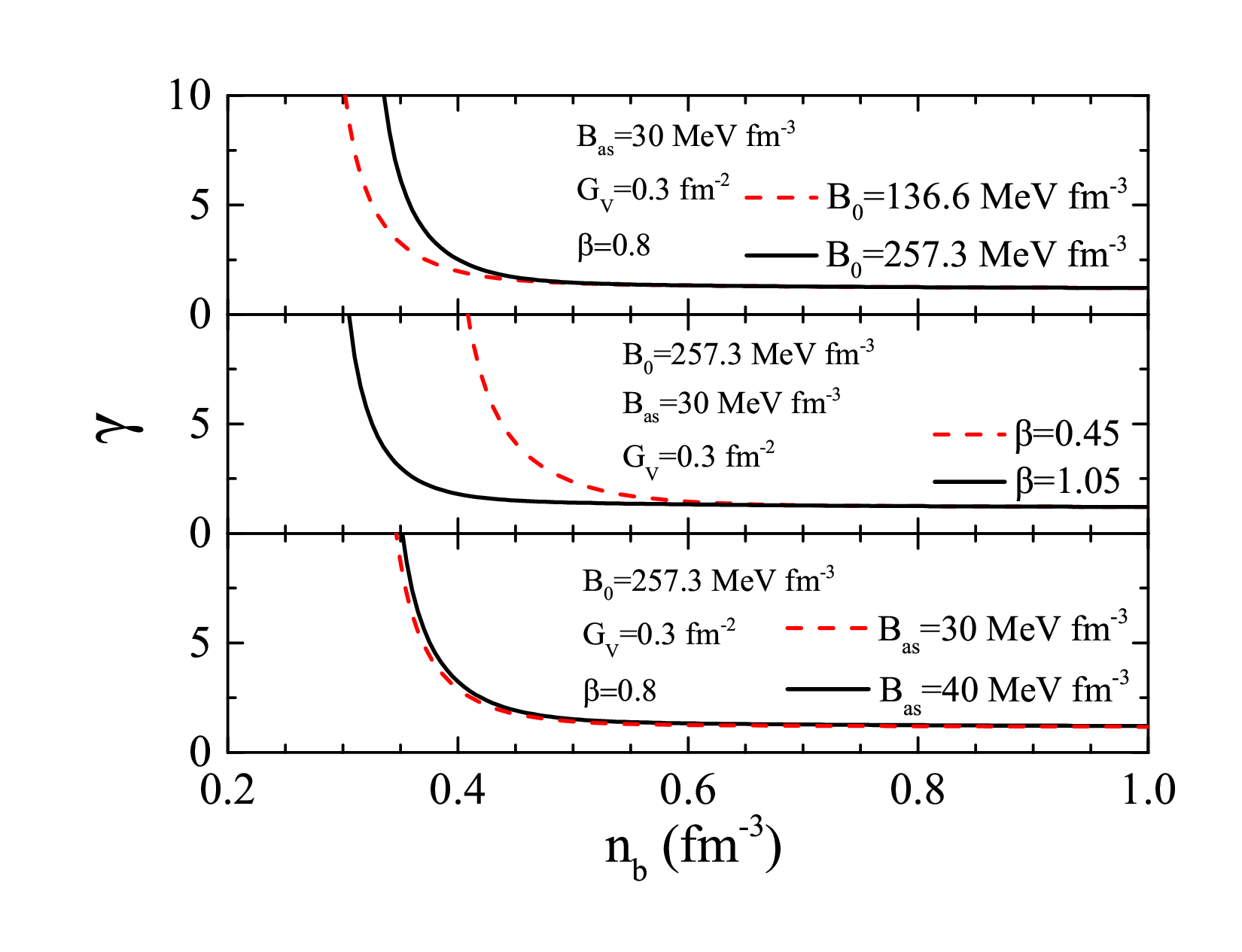}
\caption{ Similar to Fig. \ref{fig:8vs2} but for $\gamma$. }
\label{fig:9gamma}
\end{figure}

In Fig. \ref{fig:7Lambda}, we present the tidal deformability of QSs for different combinations ($B_{as}$, $\beta$) when fixing $B_0=257.3$ MeV fm$^{-3}$ and $G_{V}=0.3$ fm$^2$. 
One can find that the parameters $B_{as}$ and $\beta$ have a influence on the tidal deformability of QSs predominantly in the lower mass regime. 
This effect becomes negligible when the mass of the QS surpasses 1.6 $M_{\odot}$, which indicates the effects from the density dependence of bag pressure $B(n_b)$ mainly at lower density. 
If the canonical star faced as a QS, it would exhibit a very small tidal deformability (30 $<$ $\Lambda_{1.4}$ $<$ 100) based on the vMIT bag model, as indicated in Table~\ref{tab:QS}. This is basically outside the constraint $70<\Lambda_{1.4}<580$~\cite{LIGOScientific:2018cki}.
A similar result was also reported in Ref.~\cite{Podder:2023dey}. 

The speed of sound varying with baryon density for different combinations ($B_{as}$, $\beta$) is shown in Fig.~\ref{fig:8vs2}. 
One can see that $B_0$ and $\beta$ primarily influence the speed of sound at the initial stages of baryon density, which can be attributed to the variation of $B(n_b)$ as shown in Fig.~\ref{fig:1B}. 
Furthermore, we found a larger value of $\beta$ leads to a smaller density associated with the minimum of the speed of sound. 
The influence of $B_{as}$ = 30 and 40 MeV fm$^{-3}$ is subtle in Fig.~\ref{fig:8vs2}, although its effect becomes significant in Fig.~\ref{fig:9gamma}.
Based on the vMIT bag model with a density dependent bag pressure, the contribution to stiffness of EOS is composed of three parts: the quark degeneracy and Fermi kinetic energy, vector interaction potential and bag pressure.
At high densities, the first part tends towards the conformal limit, while the second part deviates from the conformal limit (gradually increasing). The first two parts both increase with density. 
On the other hand, the contribution of $B(n_b)$ decreases with increasing density and is significant at low densities (indeed, the change in EOS is more pronounced at low densities).
Therefore, the quark confinement, driven by a density dependent bag pressure, appears to manifest primarily as a surface effect rather than an internal interaction.

The polytropic index $\gamma$ is also used as a criterion for determining the stellar matter's inner composition. 
The $\gamma$ is mathematically expressed as
\begin{equation}
	\gamma	= \frac{\partial{\ln{P}}}{\partial{\ln{\varepsilon}}},
\end{equation}
and uses the EOS as its input.
For matter with exact conformal symmetry, matter without intrinsic scales, $\gamma$ = 1, where the energy density and pressure become proportional to each other which leads to $\gamma$ = 1.
In Fig. \ref{fig:9gamma}, we show $\gamma$ as a function of $n_b$. The graph shows a steady decrease of $\gamma$ with $n_b$ in all cases which slowly approaches the 
minimum value of 1.12.
The same results are obtained within the DDQM (Density-Dependent Quark Mass Model)~\cite{Issifu:2023qoo}.
It is observed that $\gamma$ exhibits a decreasing trend as increasing $\beta$, whereas $\gamma$ increases with higher values of $B_{as}$ and $B_0$. The Fig.~\ref{fig:5mr} and Fig.~\ref{fig:6mr} indicate that the maximum mass of QSs increases with $\beta$ while decreasing with $B_{as}$ and $B_0$, which help us conclude that the polytropic index $\gamma$ increases with the decrement of the maximum mass of QSs.

\section{Conclusions}
\label{sec:Con}
In this study, we developed the vMIT bag model by incorporating a density-dependent bag pressure where the Gaussian distribution was used to investigate the properties of QSs.
There are mainly two key parameters ($B_{as}$, $\beta$), which control the density-dependence of the bag pressure.
At low densities, the bag pressure is mainly controlled by $\beta$ while at high densities is primarily affected by $B_{as}$.
The energy per baryon and the pressure for both SQM and udQM were presented in the Fig.~\ref{fig:2PE-nb}. 
One can see that \( B_{as} \) exhibits a direct correlation with the energy per baryon but has minimal impact on the surface density, whereas \(\beta\) shows an inverse correlation with the energy per baryon.
The surface density is primarily controlled by $\beta$ and $B_0$, as illustrated in Fig.~\ref{fig:4P-nb} and Table.~\ref{tab:QS}.
Then we presented the stability window of parameters ($B_{as}$, $\beta$) in Fig.~\ref{fig:3B} based on the absolutely stable condition. We found that there is a distinct boundary of stability region, which the maximum value of $B_{as} \approx 58$ MeV fm$^{-3}$ (56 MeV fm$^{-3}$) for the case $B_0 = 257.3$ MeV fm$^{-3}$ ($B_0 = 136.6$ MeV fm$^{-3}$) with $G_V$ = 0.3 fm$^2$.
The vector interaction tends to shift the stability window to upward and to the left, while narrowing the stability window.
In Fig.~\ref{fig:5mr}, we present a series of mass-radius relations. 
There are numerous parameters ($B_{as}$, $\beta$) can simultaneously fulfill the astrophysical constraints while keeping the thermodynamic self-consistency.
Furthermore, our analysis indicates that if HESS J1731-347 is confirmed to be a QS, then 4U 1702-429, PSR J0437-4751, and PSR J0740+6620, are also likely to be QSs. 
The results show that massive QSs prefer a larger $\beta$ and smaller values of $B_{as}$ and $B_0$.
The maximum value of QSs is $\approx 2.9~M_{\odot}$ in the case $G_{V} = 0.3$ fm$^{2}$ with $B_{as}=10$ MeV fm$^{-3}$ and $\beta=0.6$. 
Furthermore, we inferred that based on the vMIT bad model with a density dependent bag pressure, it supported a massive QS heavier than $\sim 2.40~M_{\odot}$.  
To complete our analyses, we calculated the tidal deformability of QSs. We found that the canonical star (with 1.4 $M_{\odot}$), if faced as a QS, presents a very small tidal deformability ($\Lambda_{1.4} < 100$).
As we found in Fig.~\ref{fig:8vs2}, the variation of the speed of sound is mainly controlled by the density dependence of bag pressure $B(n_b)$. 
The effects of $B_0$ are mainly observed at low density ranges, and a larger value of $\beta$ leads to a smaller density associated with the minimum of the speed of sound.
Finally, we studied the polytropic index $\gamma$ and presented our results in Fig.~\ref{fig:9gamma}.
The effect of $B_{as}$ on the EOS of QSs is amplified when displayed in the context of 
$\gamma$, although it is indistinguishable in the speed of sound.
Our results show that the polytropic index of QS matter increases with the decrement of the maximum mass of QSs.

\section*{Acknowledgment}
This work was supported in part by the Fundamental Research Funds for the Central Universities under Grant (27RA2210023); the Shandong Natural Science Foundation No. ZR2023QA112, ZR2022JQ04, ZR2021QA037, ZR2019YQ01; the National Natural Science Foundation of China under Grants No. 12305148, 11975132, 12205158; Hebei Natural Science Foundation No. A2023203055.


\end{document}